\documentclass[10pt]{article}
\usepackage{sao2}
\usepackage{psfig,epsf}

\issue{2010, 54, 571--577}

\begin{document}
\markboth{M.L.~Khabibullina, O.V.~Verkhodanov, M.~Singh, A.~Pirya, N.V.~Verkhodanova, S.~Nandi}
  {RADIO SPECTRA OF GIANT RADIO GALAXIES FROM RATAN-600 DATA}
\title{Radio Spectra of Giant Radio Galaxies from RATAN-600 Data}
\author{M.L.~Khabibullina\inst{a}
O.V.~Verkhodanov\inst{a}
M.~Singh\inst{b}
A.~Pirya\inst{b}
N.V.~Verkhodanova\inst{a}
S.~Nandi\inst{b}
}
\institute{
$^a$\saoname; \\ $^b$\ARIES}

\date{November, 2009}{Jule, 2010}
\maketitle

\begin{abstract}
Abstract.Measurements of the flux densities of the extended components of seven giant radio galaxies 
obtained using the RATAN-600 radio telescope at wavelengths of 6.25 and 13 cm are presented. The 
spectra of components of these radio galaxies are constructed using these new RATAN-600 data together 
with data from the WENSS, NVSS, and GB6 surveys. The spectral indices in the studied frequency range 
are calculated, and the need for detailed estimates of the integrated contribution of such objects to the 
background emission is demonstrated.\\

\end{abstract}
\maketitle

\section{INTRODUCTION}
Giant radio galaxies (GRGs) are the largest radio 
sources in the Universe, with linear sizes of more 
than 1 Mpc. Studies of these objects began with 
3C 236 \cite{strom}. They display primarily morphological type
FR II \cite{fr}, and are identified with giant elliptical galaxies
 and quasars. GRGs are rare compared to ordinary
 radio galaxies, hindering statistical studies and 
detailed studies of their formation as a population. 
Several groups are undertaking studies of their properties,
 aimed at explaining their huge sizes \cite{shom_mack} - \cite{mach}.
As was noted by Jamrozy et al. \cite{jamr_mach}, the influence
of GRGs on the surrounding environment could be 
more extensive than for ordinary radio galaxies, whose 
sizes are an order of magnitude smaller. It is an 
important fact that GRGs have sizes that are comparable
 to or larger than the dimensions of groups 
of galaxies. Therefore, they are believed to play an 
important role in the formation of the large-scale 
structure of the Universe \cite{jamr_mach}.
One possible explanation for the large sizes of 
GRGs is that they are an effect of orientation. However,
 the idea that these sizes could represent a maximum
 projection onto the plane of the sky, compared 
to ordinary radio galaxies, is not confirmed by observations
 with the Effelsberg telescope \cite{shom_bruyn}. Analyzing
 various properties of radio sources (asymmetry, 
energy densities in the lobes, variation in the lobe 
pressure with redshift, spectral age, ambient density), 
Schoenmakers et al. \cite{shom_bruyn} showed that the asymmetry of
the arm lengths for GRGs is somewhat higher than 
for 3CR radio galaxies having smaller dimensions, 
which is difficult to explain with orientation effects. 

There are currently two main views about the origin
 of the large sizes of GRGs: (1) they are related to 
properties of the energy released by the jets emerging
 from their nuclei; (2) they are determined by the 
properties of the ambient medium in the local group 
of galaxies. Schoenmakers et al. \cite{shom_bruyn} suggest that
the latter effect is most important. They concluded 
that GRGs may represent the oldest, comparatively 
powerful members of a group of radio sources, whose 
radio emission has continued to develop. Their spectral
 analysis indicated that the lobes of GRGs have 
higher pressures than the ambient medium, and that 
the redshift dependence of the lobe pressures could 
also be due to selection effects. The conclusions of \cite{shom_bruyn}
are supplemented by the analysis of radio and optical 
(SDSS, APM) data for radio galaxies and quasars 
carried out by Komberg and Pashchenko \cite{komberg}, who
found that, in addition to the influence of the surrounding
 medium, the sizes of giant radio sources 
can be explained by the presence of a population of 
long-lived, radio-loud active nuclei, which, in turn,
can evolve into GRGs.

It is also of interest to investigate how various
characteristics of GRGs (size,shape,orientation)
affect maps of the microwave background radiation
\cite{verkh_kh}. Although their contribution to maps of
the millimeter background are modest, their angular
sizes (to 10\arcmin) create problems in connection with
distinguishing components due to spectral-index
variations at the locations of the extended radio lobes
of these galaxies. Therefore, it is of interest to estimate
and take into account the possible contribution of
GRGs to anisotropy of the background radiation, due
both to their millimeter radiation and to effects arising We present here the first RATAN-600 measure-
during the identification of components on the scales ments of the centimeter-and decimeter-wavelength
of multipoles $\ell\ge500$
in various frequency ranges. flux densities of seven giant radio galaxies. Note that
radio galaxies of arcmin size had been observed earlier
by Soboleva with RATAN-600 \cite{soboleva} in cm/mm
wavelength range. It was detected that morphological
structures had approximately same spectral indices.

\section{RATAN-600 data}
\subsection{Observations with RATAN-600}

The observations of the GRGs were carried out
on the northern sector of the RATAN-600 in mid-
December 2008, using continuous-spectrum radiometers
 at the primary feed, operating at wavelengths
 of 1.38, 2.7, 3.9, 6.25, 13, and 31\,cm cite{nizhelsky}.
Unfortunately, due to the presence of various types
of interference during the observing period, only the
data at 6.25 and 13\,cm were suitable for our analysis.
The dimensions of the antenna beam in the central
section at the observed elevations were 43\arcsec
and 90\arcsec.
Fortunately, both wavelengths had spectral-analyzer
subchannels, making it possible to effectively combat
interference. The subchannels at 31\,cm proved not
as useful in this sense. Due to the short observing
times, the required sensitivity ($<$20 mK) was not
achieved at wavelengths shorter than 6\,cm. A list of
the observed sources is given in Table\,1, and a journal
of the observations in Table\,2. Despite the fact that
the GRG\,1343+3758 was observed twice, it was not
possibleto achieveasufficient signal-to-noise ratio
to detect this source.

\begin{table*}[!tbp]
\begin{center}
\caption{Main parameters of observed giant radio galaxies}
\begin{tabular}{|ccccrr|}
\hline  source        & coordinates     & redshift&type & angular    &flux      \\
Source name           &Coordinates      &Redshift &Type&Angular size,&Flux density  \\
		      & RA/Dec (J2000.0)&         &    &arcmin       &(1.4 GHz), mJy \\
\hline
GRG 0139+3957 & 013930+395703   &  0.211  & II  &  5.7    &  801.1   \\
GRG 0912+3510 & 091252+351016   &  0.249  & II  &  6.2    &  157.4   \\
GRG 1032+2759 & 103214+275600   &  0.085  & II  &  11.0   &  284.1   \\
GRG 1343+3758 & 134255+375819   &  0.227  & II  &  11.3   &  131.0   \\
GRG 1400+3017 & 140040+301700   &  0.206  & II  &  10.8   &  451.9   \\
GRG 1453+3308 & 145303+330841   &  0.249  & II  &  5.7    &  455.5   \\
GRG 1552+2005 & 155209+200524   &  0.089  & II  &  19.6   &  2385.6  \\
GRG 1738+3733 & 173821+373333   &  0.156  & II  &  6.5    &  236.0   \\
\hline
\end{tabular}
\end{center}
\end{table*}

\begin{table*}[!tbp]
\begin{center}
\caption{Observed regions of the GRGs.
Cross-sections: 'c' means central,
'n' means Northern, 's' means southern. $N_t$ is the number of passes.
Coordinates (right ascension+ declination) of component centers are given
at the epoch of J2000.0.}
\begin{tabular}{|cccc|}
\hline
Source name      &section &coordinates of    &Number of transits $N_t$ \\
		 &        &center of observed&                        \\
		 &        &region            &                         \\
\hline
GRG 0139+3957  & c     &    013927.4+395653 &  1    \\
\hline
GRG 0912+3510  & n     &    091252.0+351231 &  5    \\
GRG 0912+3510  & s     &    091250.0+350631 &  1    \\
\hline
GRG 1032+2756  & n     &    103212.5+275925 &  3    \\
GRG 1032+2756  & c     &    103214.4+275555 &  3    \\
GRG 1032+2756  & s     &    103215.1+275115 &  1    \\
\hline
GRG 1343+3758  & c     &    134255.0+375819 &  2    \\
\hline
GRG 1400+3017  & n     &    140045.0+302214 &  3    \\
GRG 1400+3017  & s     &    140038.4+301325 &  3    \\
\hline
GRG 1453+3308  & n     &    145302.0+331046 &  4    \\
GRG 1453+3308  & c     &    145303.0+330856 &  2    \\
GRG 1453+3308  & s     &    145301.4+330556 &  1    \\
\hline
GRG 1552+2005  & c     &    155209.0+200524 &  8    \\
\hline
GRG 1738+3733  & n     &    173820.6+373658 &  2    \\
GRG 1738+3733  & c     &    173821.0+373333 &  2    \\
GRG 1738+3733  & s     &    173821.8+373108 &  1    \\
\hline
\end{tabular}
\end{center}
\end{table*}

Depending on the position angle of the radio
structure, from one to three cross-sections through
the sourcewere made (Table\,2). Thenumber of
transits of the objects through the telescope beam
was limited by the total observing time granted for
the project.

\subsection{Reduction}

To tie the flux densities to the international scale
\cite{baars}, we carried out observations of calibrator sources
from the RATAN-600 standard list \cite{trushk,verkh_erukh}. The
transit curves for the sources were analyzed in the
FADPS standard reduction system \cite{verkh_erukh,verkh}. When
analyzing each component in the recordings, we
subtracted a low-frequency trend obtained with an
8\arcmin
smoothing window. We estimated the flux densities
using the integrals beneath the curves for the transits
 of the sources through the RATAN-600 beam,
approximated by sets of Gaussians. The noise levels
in therecordingsfor single transits at an elevation
of $76^\circ$
were 8.1, 5, 36, 3.3, and 65 mK/$s^{1/2}$ at 1.38,
2.7, 3.9, 6.25, and 13 cm, respectively. The 6.25 and
13-cm flux densities are presented in Table\,3, together
with the integrated flux densities in the measured
components (marked in Fig.\,1) calculated from the

\begin{table*}[!tbp]
\begin{center}
\caption{Flux densities of the components (in mJy) according
to the RATAN-600, WENSS, NVSS, and GB6}
\begin{tabular}{|r|rrrrr|}
\hline
Source      &6.25\,cm &13\,cm& 92\,cm  &21\,cm & 6.25\,cm  \\
Component   &RATAN    &RATAN   &  WENSS  &  NVSS  &   GB6    \\
\hline
0139+3957w  &    470    &   857   &    --   &  2120  &   656    \\
	c  &    139    &   252   &    --   &   744  &   317    \\
	e  &    82     &   212   &    --   &   133  &          \\
0912+3510n  &    30     & $<$120  &    160  &    56  & $<$20    \\
	s  &    73     &   166   &    512  &   101  &    20    \\
1032+2759n  &    66     & $<$120  &    --   &    92  & $<$20    \\
	c  &    52     & $<$120  &    --   &    75  &    59    \\
	s  &   106     &   148   &    --   &   138  &    56    \\
1400+3017n  &    63     &   212   &   1258  &   333  &    73    \\
	s  &    46     &   194   &   1053  &   155  &    37    \\
1453+3308n  &    85     &   109   &    420  &   245  & $<$20    \\
	c  &   123     &   226   &    593  &   149  &   131    \\
	s  &   109     &   116   &    488  &    89  & $<$20    \\
1552+2005w  &    82     &   212   &    --   &   133  & $<$20    \\
	e  &   139     &   252   &    --   &   744  &   317    \\
	ee &   470     &   857   &    --   &  2120  &   656    \\
1738+3733n  &    54     &   160   &    152  &    64  & $<$20    \\
	c  &    63     &   174   &    720  &   117  &    93    \\
	s  &    50     &   123   &    133  &    58  & $<$20    \\
\hline
\end{tabular}
\end{center}
\end{table*}

\begin{figure*}
\centerline{
\vbox{
\hbox{
 \psfig{figure=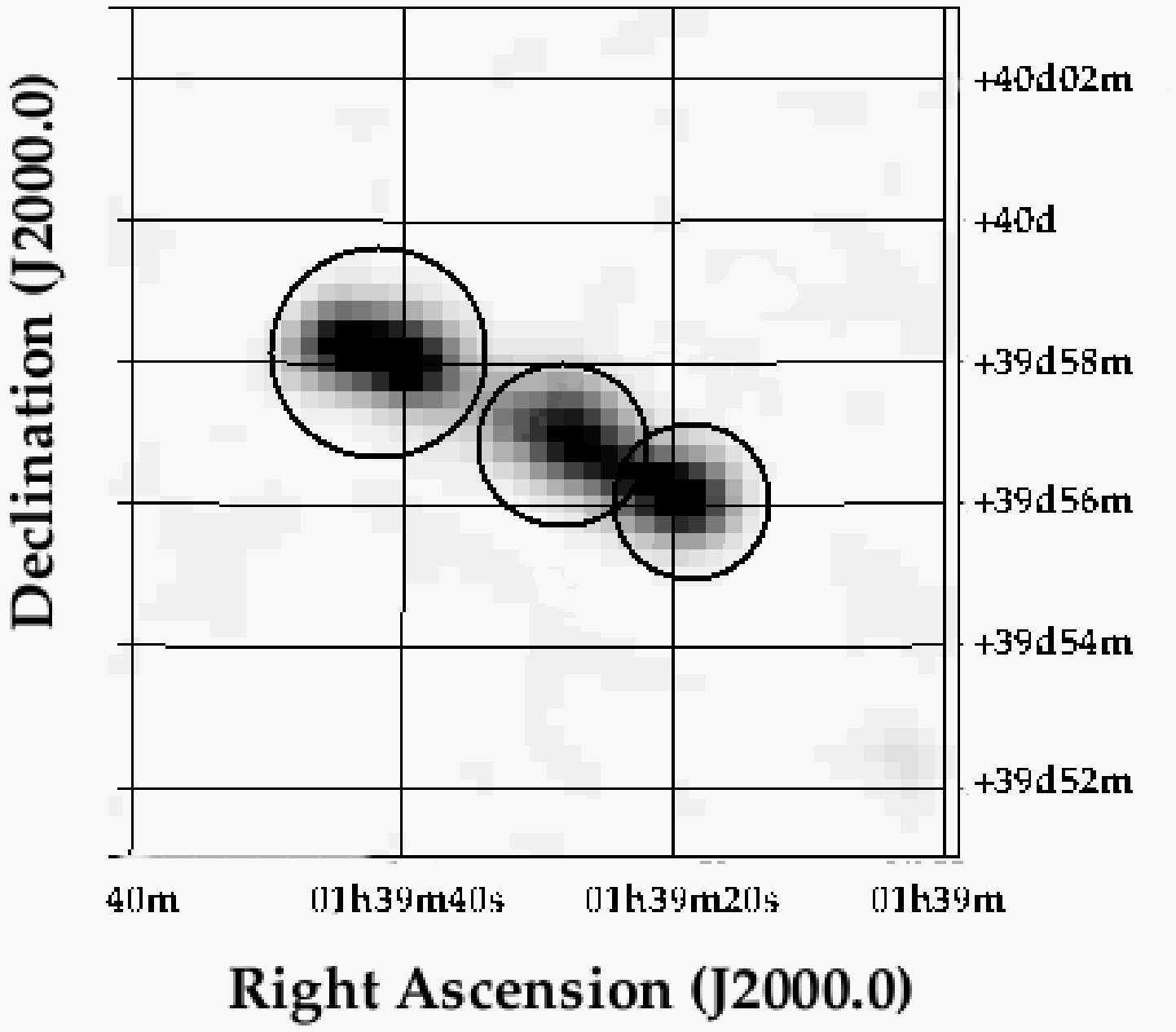,width=6.5cm}
 \psfig{figure=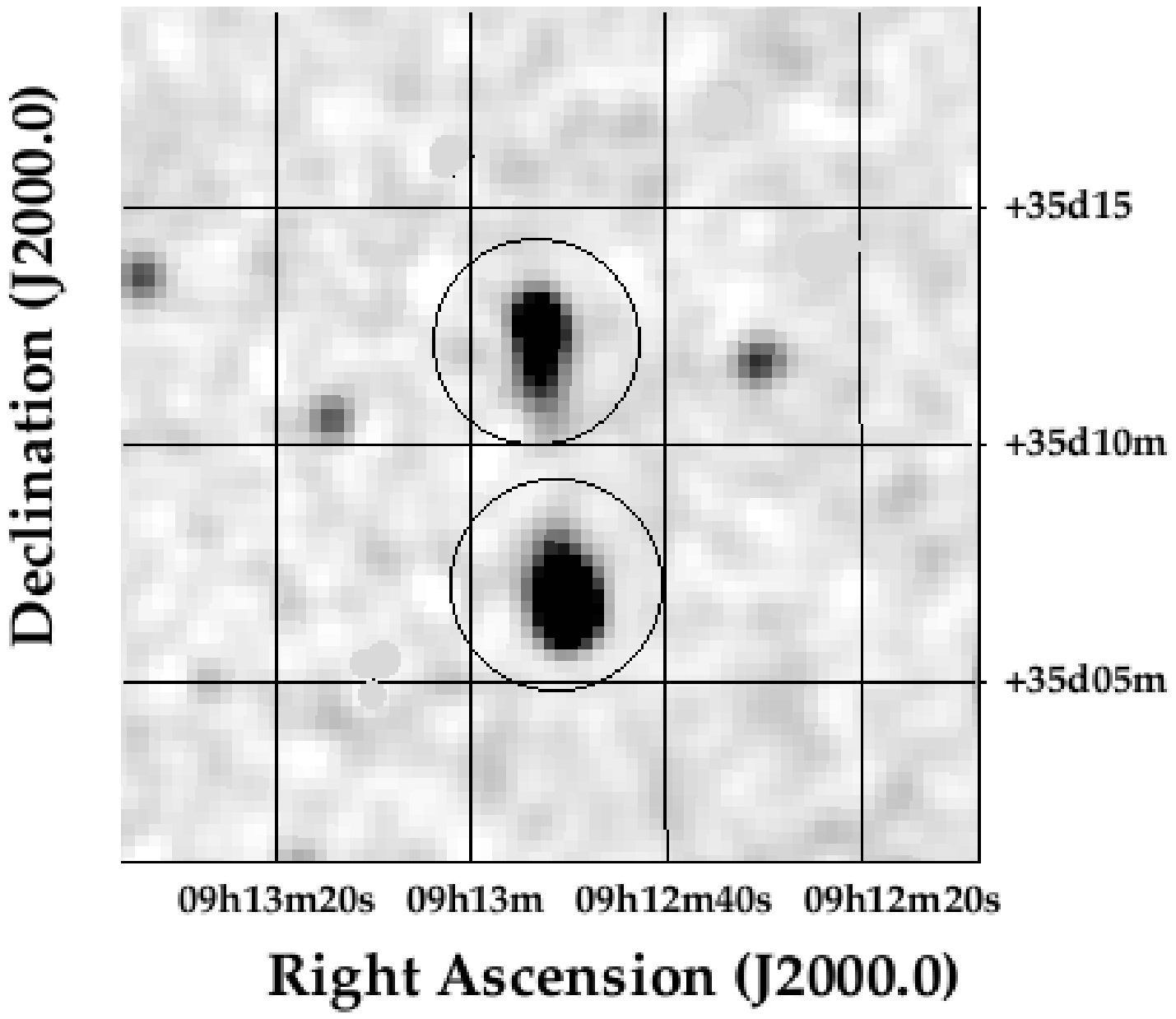,width=6.5cm}
}
\hbox{
 \psfig{figure=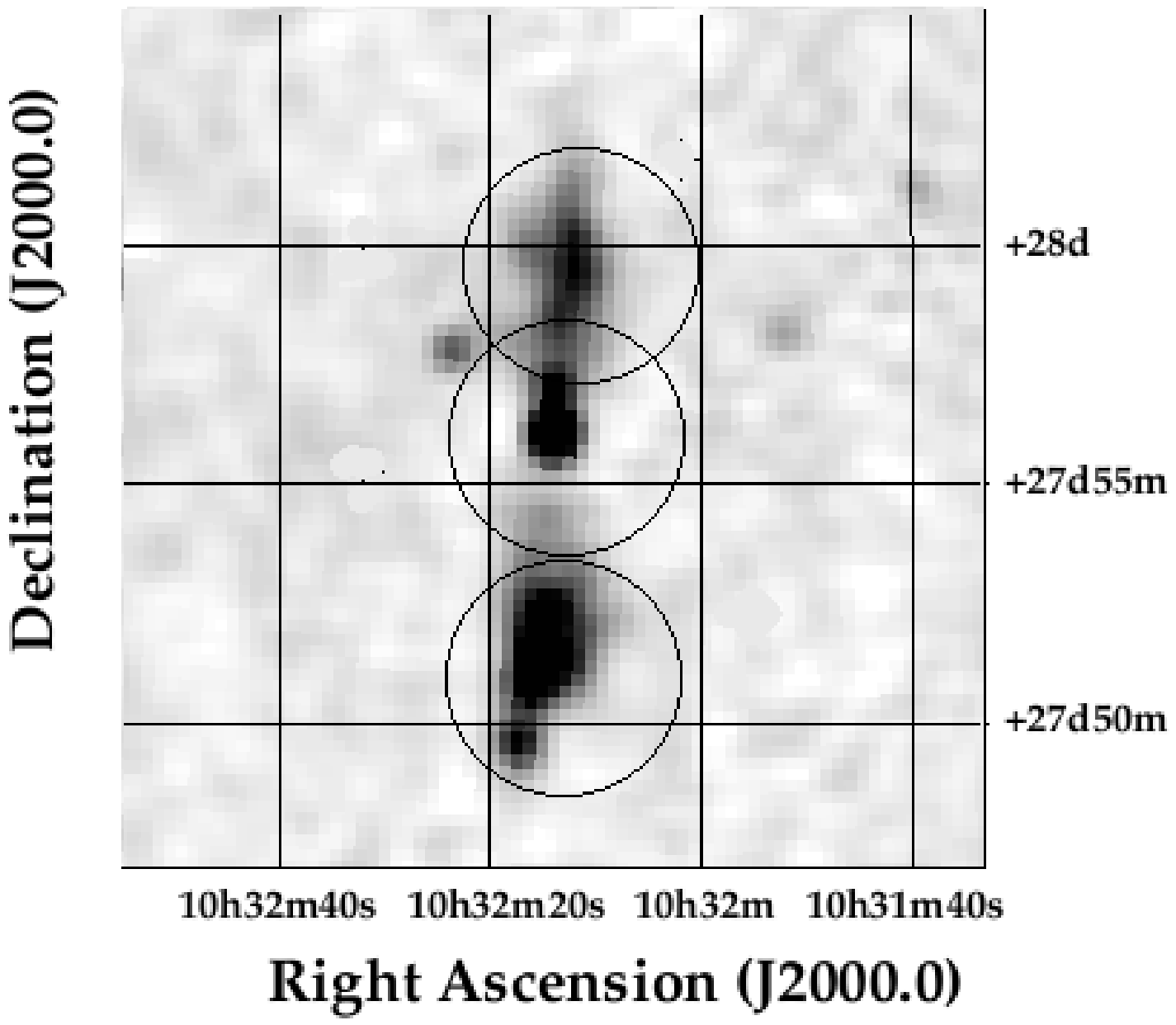,width=6.5cm}
 \psfig{figure=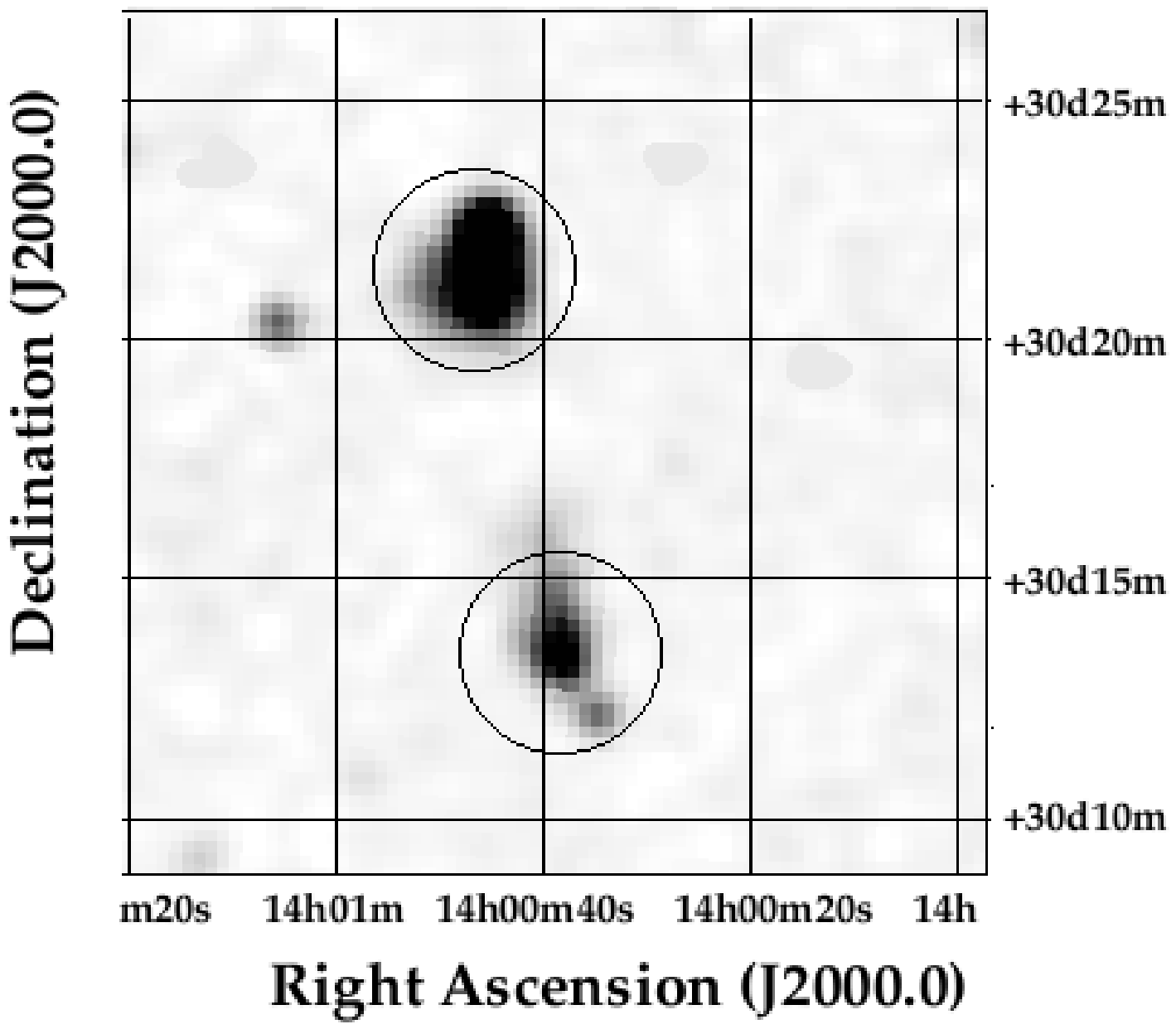,width=6.5cm}
}
\hbox{
 \psfig{figure=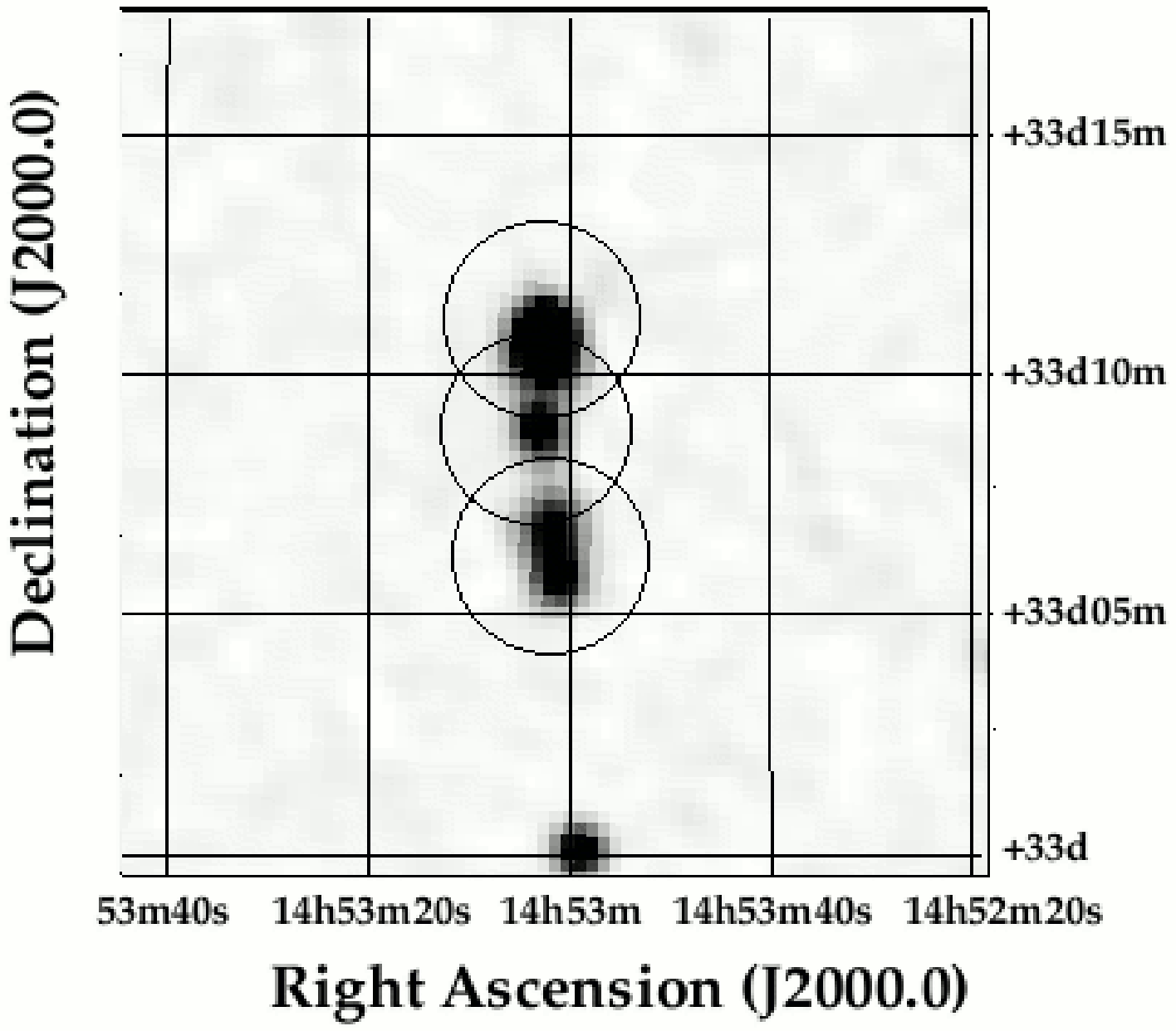,width=6.5cm}
 \psfig{figure=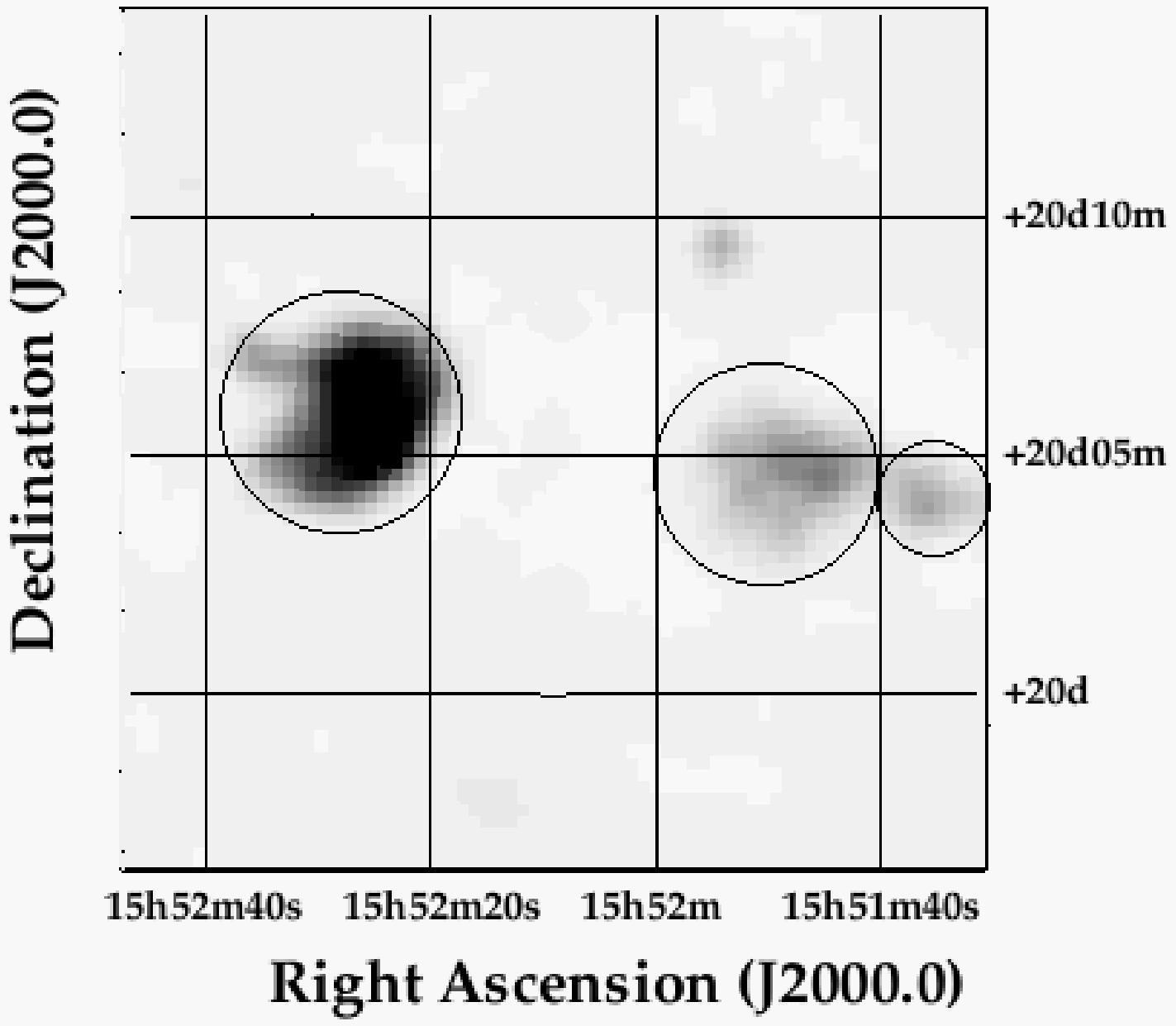,width=6.5cm}
}
\hbox{
 \psfig{figure=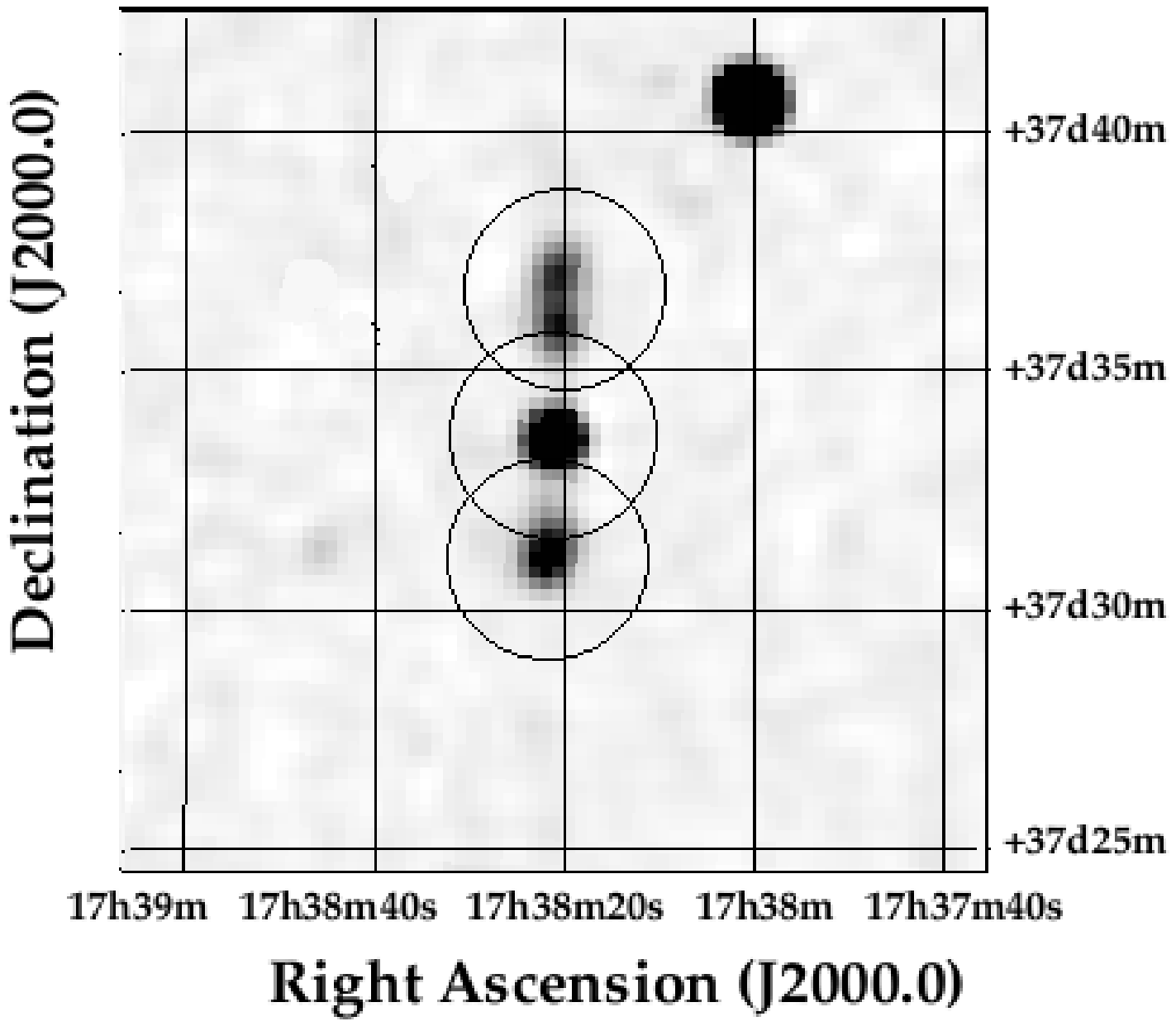,width=6.5cm}
 \mbox{\hspace*{8cm}}
}}}
\caption{Radio images of the GRGs from the NVSS survey. The circles mark
the components observed on the RATAN-600.}
\end{figure*}

NRAO VLA Sky Survey (NVSS) maps \cite{condon_cotton}, Westerbork
 Northern Sky Survey (WENSS) maps \cite{rengel},
and data from the Green Bank GB6 catalog \cite{greg}. We
also used the CATS database \cite{verkh_trushk} to identify objects
and estimate the parameters of components.

The uncertainties in the 6.25-cm RATAN-600 flux
densities for sources with flux densities $>$50\,mJy are
$\sim$10\%,and for sources with flux densities $<$50\,mJy
13\% at wavelength of 6.25\,cm. The undertainties
at 13\,cm are 10\% for sources with flux densities
$>$180\,mJy and 15\% for sources with flux densities
$<$180\,mJy.

\subsection{Spectra}

We constructed spectra for the radio components
using the data in Table\,3. We described the spectra
using the parametric formula $ lg\,S(\nu) = A + Bx + C f(x)$,
where $S$ s the flux density in Jy, $x$ the logarithm
 of the frequency $\nu$ in MHz, and $f(x)$ one of
the following functions: $exp(-x)$, $exp(x)$, or $x^2$. We
used the '{\it spg}'. system to analyze the spectra \cite{verkh2}.
The spectra of the components are shown in Fig.\,2.
Analytical fits of the curves representing the continuous
 spectra of the GRG components are presented in
Table\,4.

\begin{figure*}
\centerline{
\vbox{
\hbox{
\psfig{figure=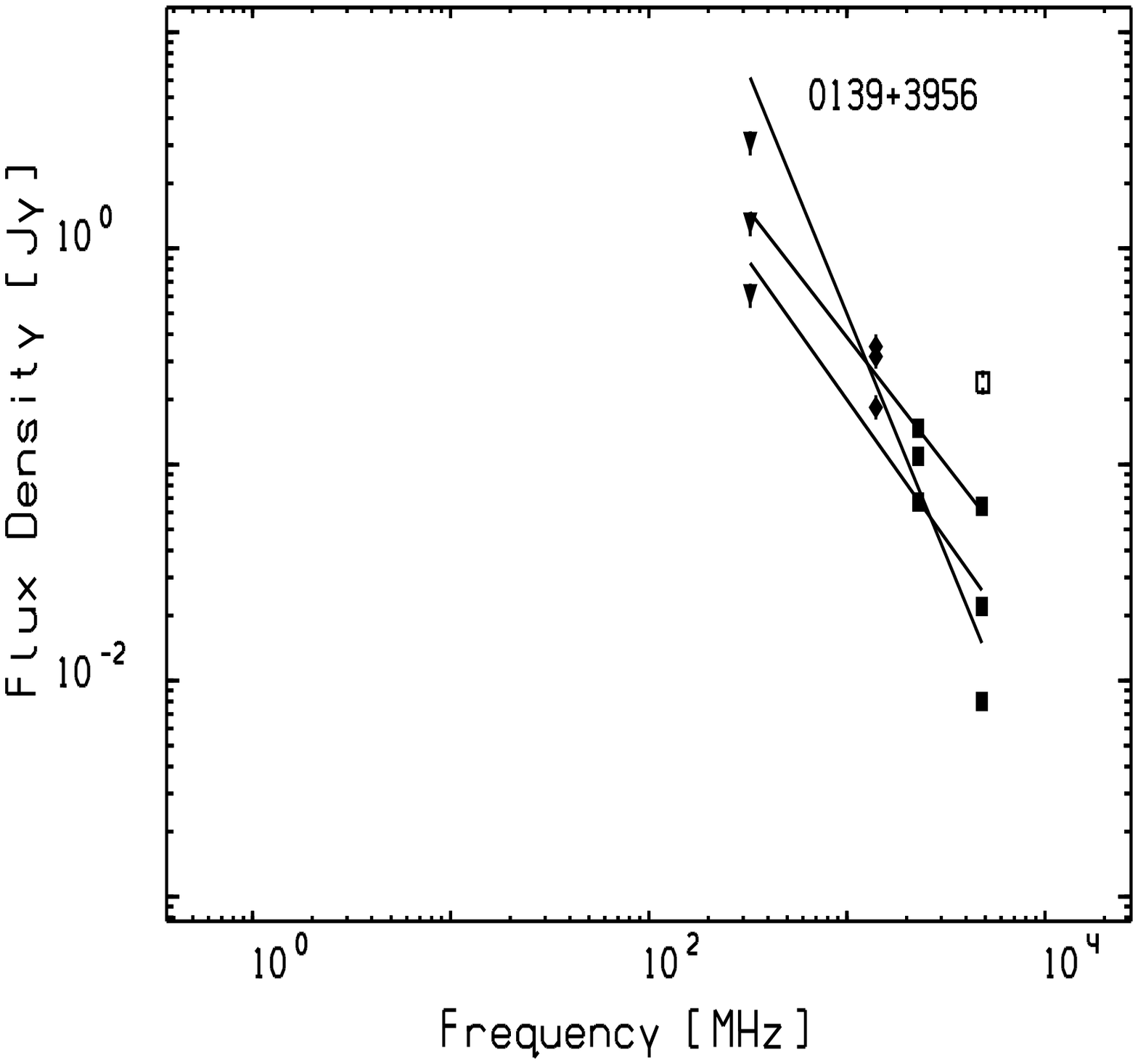,width=5cm}
\psfig{figure=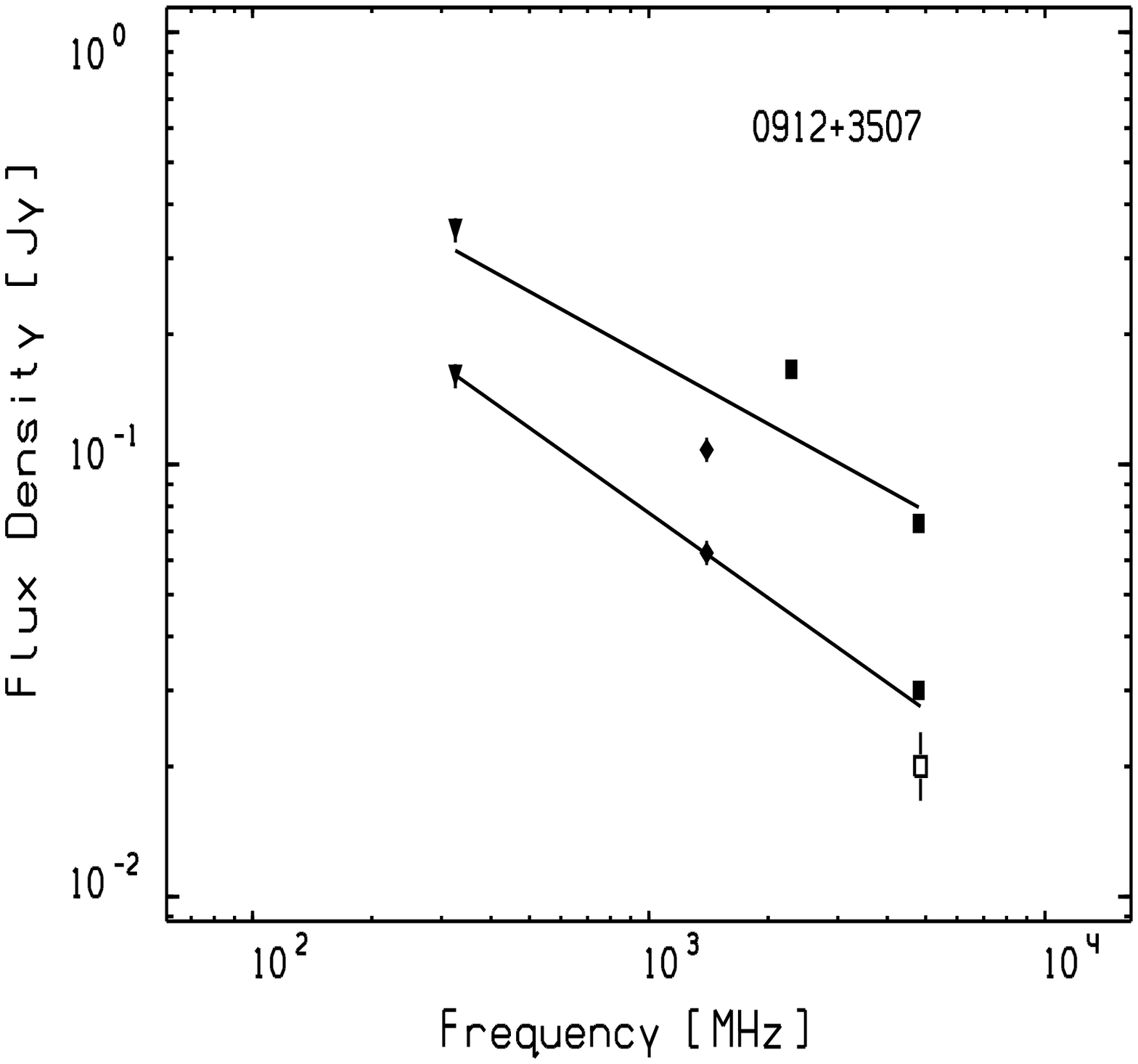,width=5cm}
\psfig{figure=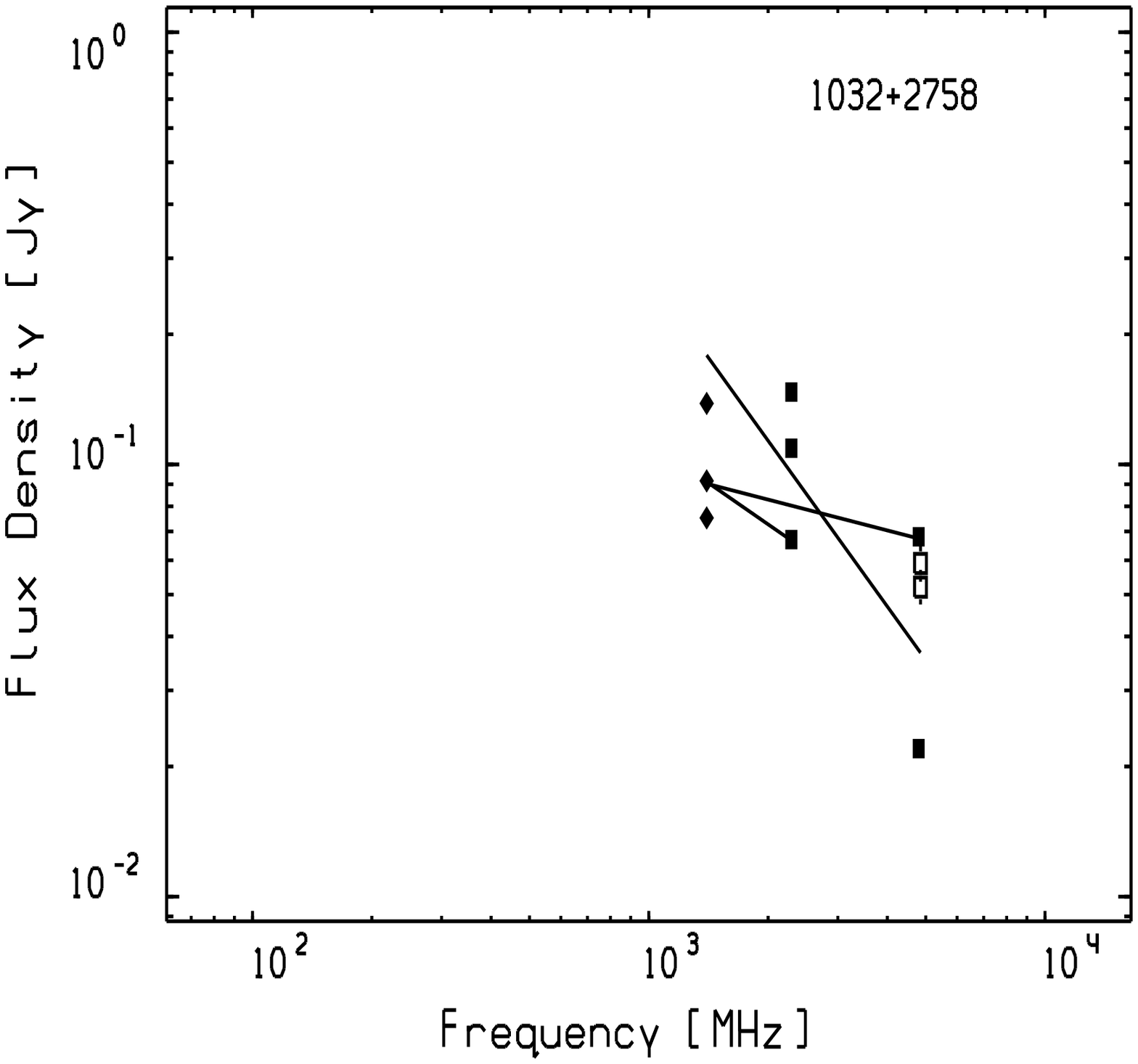,width=5cm}
}
\hbox{
\psfig{figure=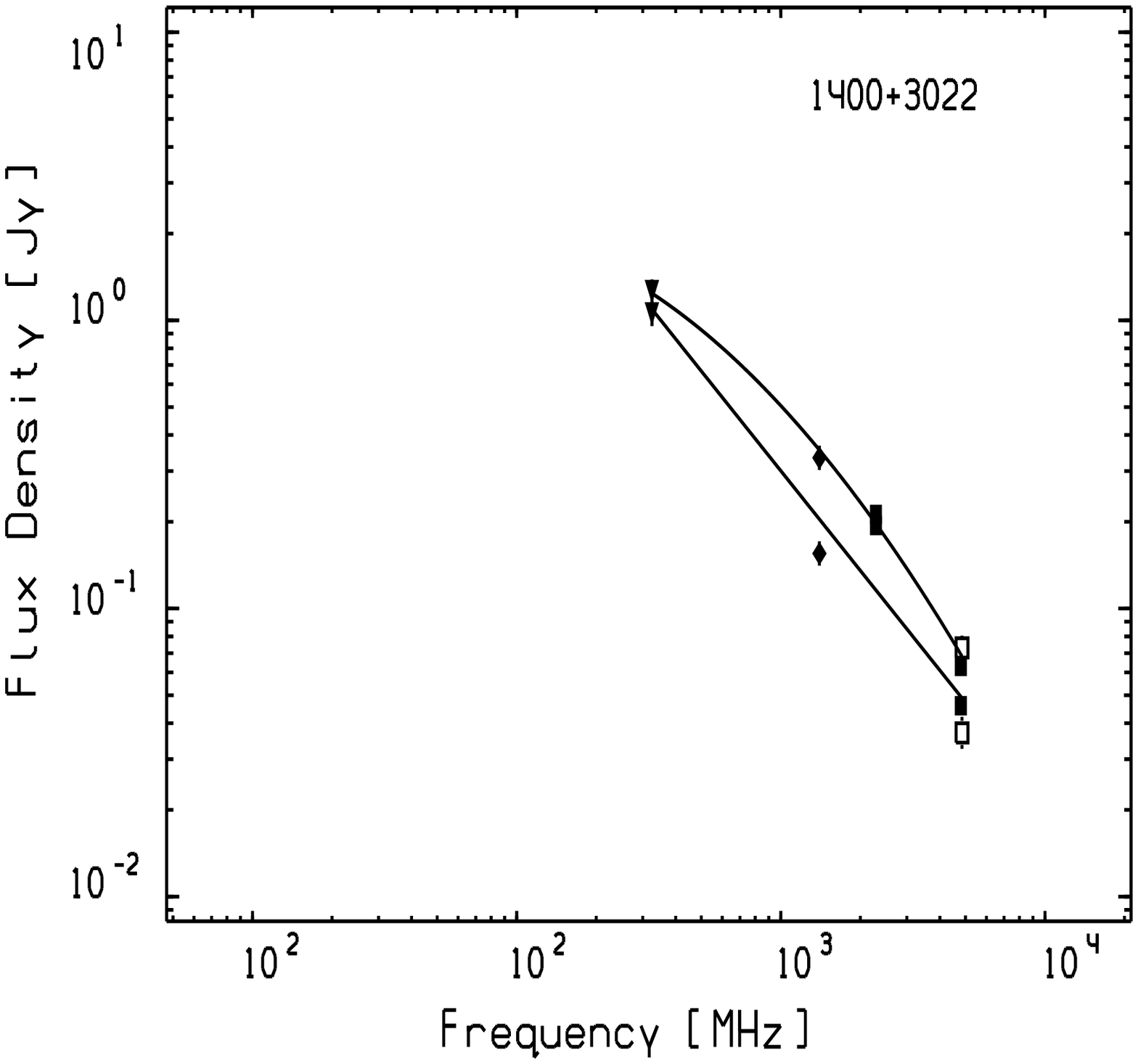,width=5cm}
\psfig{figure=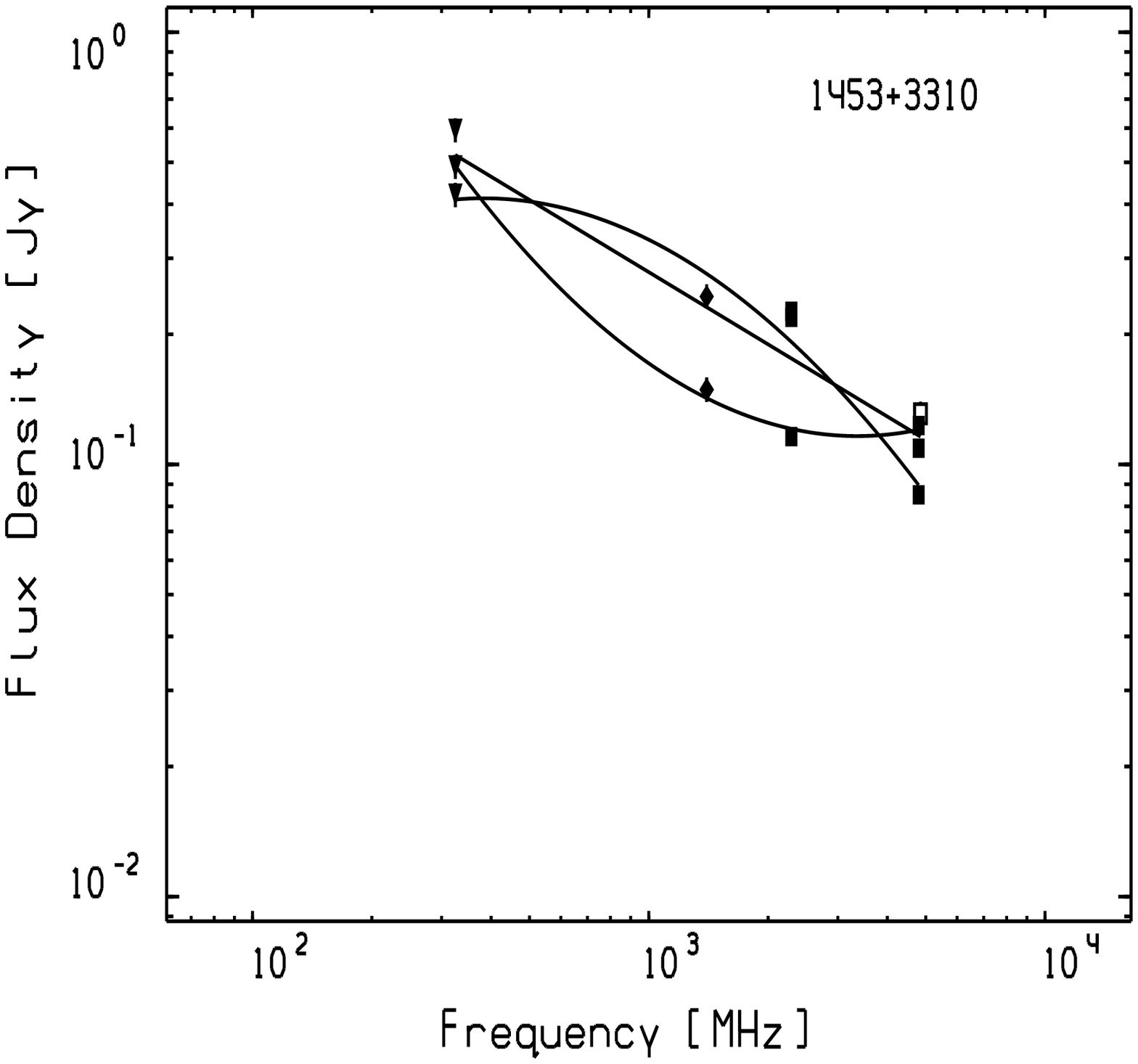,width=5cm}
\psfig{figure=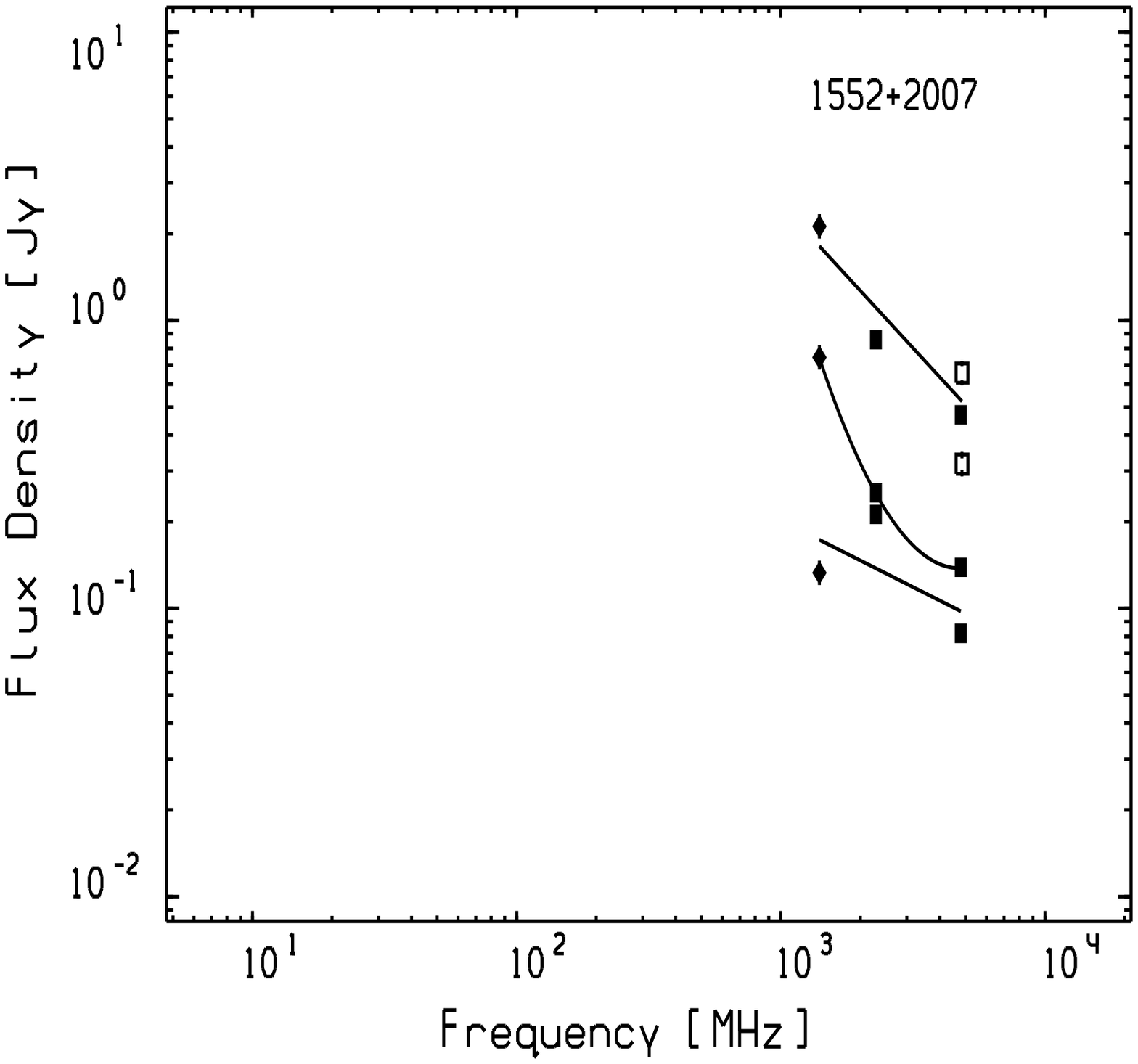,width=5cm}
}
\hbox{
\psfig{figure=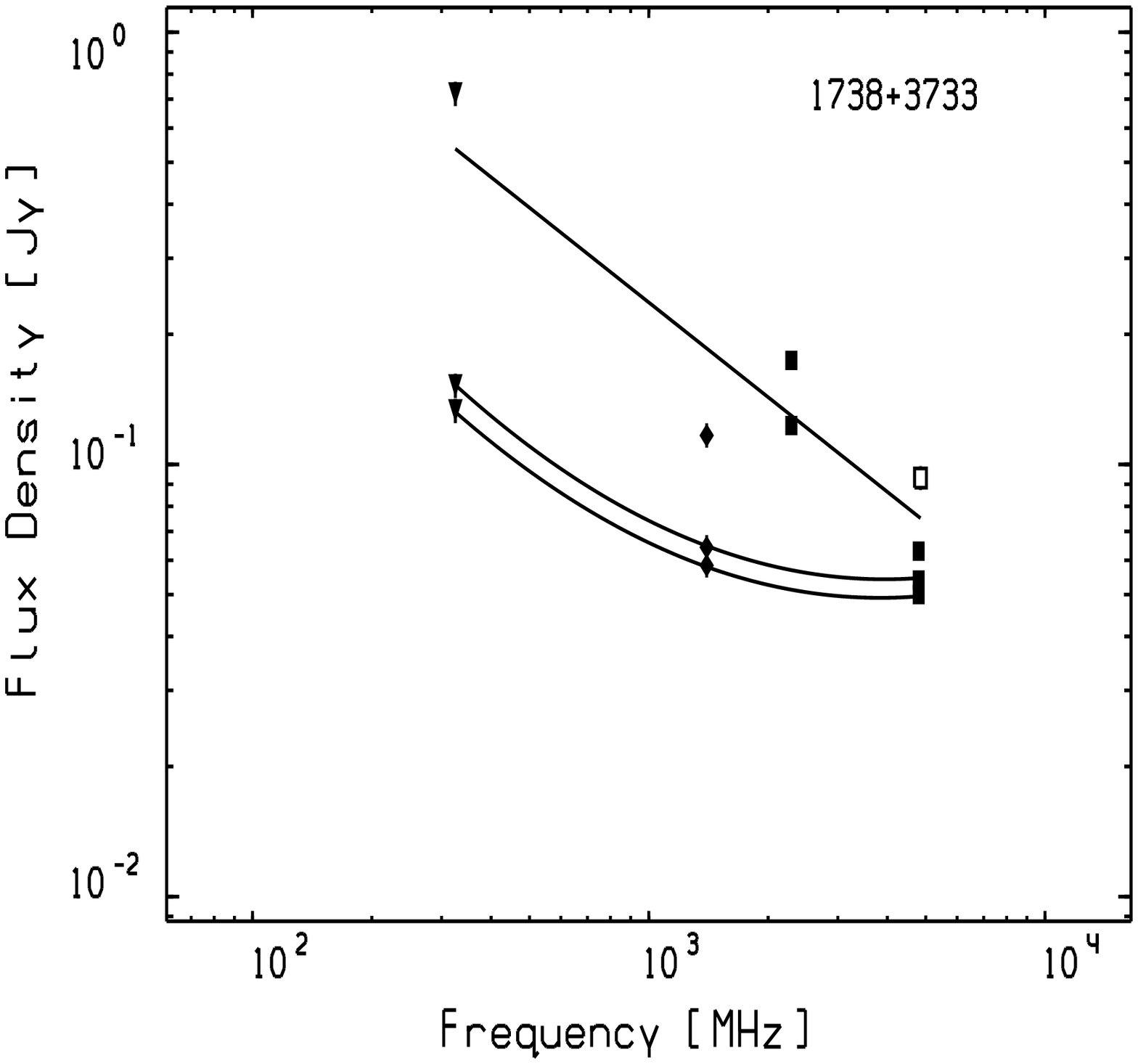,width=5cm}
}
}}
\caption{Radio spectra of the GRG components constructed using the
RATAN-600, GB6, NVSS, and WENSS data (Table\,3).}
\end{figure*}

\begin{table}[!tbp]
\begin{center}
\caption{Approximate dependences for the continuous
radio spectra of the GRG components from 92 to 6.25\,cm}
\begin{tabular}{|rr|}
\hline
Source      &  Radio spectrum \\
component   &                 \\
\hline
0139+3957w  &  $ 3.182-1.294x          $ \\
	c  &  $ 6.409-2.237x          $ \\
	e  &  $ 3.135-1.182x          $ \\
0912+3510n  &  $ 0.847-0.653x          $ \\
	s  &  $ 0.771-0.508x          $ \\
1032+2759n  &  $ 3.264-1.275x          $ \\
	c  &  $-0.299-0.237x          $ \\
	s  &  $ 0.961-0.636x          $ \\
1400+3017n  &  $-0.914+1.406x-0.400x^2 $ \\
	s  &  $ 2.928-1.150x          $ \\
1453+3308n  &  $-4.053+2.841x-0.550x^2 $ \\
	c  &  $ 1.110-0.555x          $ \\
	s  &  $ 6.626-4.288x+0.608x^2 $ \\
1552+2005w  &  $ 0.695-0.463x          $ \\
	e  &  $34.372-19.175x+2.609x^2$ \\
	ee &  $ 3.394-0.997x          $ \\
1738+3733n  &  $ 3.496-2.681x+0.374x^2 $ \\
	c  &  $ 1.559-0.728x          $ \\
	s  &  $ 3.685-2.754x+0.383x^2 $ \\
\hline
\end{tabular}
\end{center}
\end{table}

\section{DISCUSSION}

The constructed GRG spectra (Fig\,2) display a
variety of properties. It is obvious that the spectral
indices and spectral shapes of the observed GRG

components differ appreciably, from very steep spectra,
 as in GRG\,0139+3957, to fairly flat spectra, as in
GRG\,1738+3733.

For each component of the studied GRGs, we
calculated the spectral indices as the tangent of the
slope of the curve approximating the spectrum at the
given frequency. The results for 6.25 and 13\,cm are
presented in Table\,5.

GRG\,1738+3733 stands out among the observed
sources, due to the similarity of the radio spectra and
spectral indices for both its extended components.

Note that the variation of the spectral indices of
GRGs with distance from the center of the galaxy was
noted earlier in \cite{shom_bruyn}. This is due to the variation of the
particle energy in the components with distance from
the host galaxy due to variations in the pressure of the
gas flowing around the lobes, i.e., the pressure of the
surrounding medium.

\begin{table}[!tbp]
\begin{center}
\caption{Spectral indices for the GRG components at 6.25 and 13\,cm}
\begin{tabular}{|r|cc|}
\hline
Source         &      Spectral index     \\
component      & 6.25\,cm    & 13\,cm    \\
\hline
0139+3957w  &      -1.29  &  -1.29       \\
	c  &      -2.24  &  -2.24       \\
	e  &      -1.18  &  -1.18       \\
0912+3510n  &      -0.65  &  -0.65       \\
	s  &      -0.51  &  -0.51       \\
1032+2759n  &      -1.27  &  -1.27       \\
	c  &      -0.24  &  -0.24       \\
	s  &      -0.64  &  -0.64       \\
1400+3017n  &      -1.54  &  -1.28       \\
	s  &      -1.15  &  -1.15       \\
1453+3308n  &      -1.21  &  -0.86       \\
	c  &      -0.56  &  -0.56       \\
	s  &       0.19  &  -0.20       \\
1552+2005w  &      -0.46  &  -0.46       \\
	e  &       0.03  &  -1.63       \\
	ee &      -1.00  &  -1.00       \\
1738+3733n  &       0.07  &  -0.17       \\
	c  &      -0.73  &  -0.73       \\
	s  &       0.07  &  -0.18       \\
\hline
\end{tabular}
\end{center}
\end{table}

Note also that the RATAN-600 observations
have made it possible to refine the spectra of the
GRG components and, via extrapolation of the radio
spectra, estimate their fluxes at millimeter wavelengths.
 These extrapolated flux densities exceed
0.6\,mJy. With an expected total number of GRGs
of several hundred \cite{verkh_kh}, their contribution to the
background radiation can, in principle, lead to a
bias in calculations of the level of fluctuations of the
background, as well as problems in distinguishing
the background signal. To estimate this bias, we have
begun composing a catalog of GRGs with modest
flux densities ($<$100\,mJy) over the entire sky.

\noindent
{\small
{\bf Acknowledgments}.

This work has made use of the CATS database
of radio-astronomy catalogs \footnote{\tt http://cats.sao.ru} \cite{verkh_trushk}
and the FADPS
radio-astronomy data-reduction system
\footnote{\tt http://sed.sao.ru/$\sim$vo/fadps\_e.html} \cite{verkh}. This
work was supported by the Program of State Support
for Leading Scientific Schools of the Russian Federation
 (the school led by S.\,M.\,Khaikin) and the Russian
Foundation for Basic Research (project no. 09-0292659-
IND). O.V.V. also thanks the Foundation for
the Support of Russian Science for partial support
(a Doctors of Science of the Russian Academy of
Sciences grant) and the ``Dynasty'' Foundation.


\begin{thebibliography}{}


\bibitem{strom}
[1]R.~G.~Strom,  A.~G.~Willis, Astron. Astrophys. {\bf 85}, 36  (1980).


\bibitem{fr}
[2]B.~L.~Fanaroff,  J.~M.~Riley, \mnras {\bf 167}, 31p (1974).


\bibitem{shom_mack}
[3]A.~P.~Schoenmakers, K.-H.~Mack,  A.~G.~de Bruyn, et al.,
   Astron. Astrophys. Suppl. Ser. {\bf 146}, 293 (2000).


\bibitem{shom_bruyn}
[4]A.~P.~Schoenmakers,  A.~G.~de Bruyn, H.~J.~A.~Roettgering, H.~ van der Laan,
   Astron. Astrophys. {\bf 374}, 861 (2001).


\bibitem{lara_marq}
[5]L.~Lara, I.~Marquez, W.~D.~Cotton, et al.,
  Astron. Astrophys. {\bf 378}, 826 (2001).


\bibitem{lara_giov}
[6]L.~Lara, G.~Giovannini, W.~D.~Cotton, et al.,
  Astron. Astrophys. {\bf 421}, 899  (2004).


\bibitem{saripalli}
[7]L.~Saripalli,  R.~W.~Hunstead, R.~Subrahmanyan, E.~Boyce,
    Astron. J. {\bf 130}, 896 (2005).


\bibitem{konar_saikia}
[8]C.~Konar, D.~J.~Saikia,  C.~H.~Ishwara-Chandra,  V.~K.~Kulkarni,
 MNRAS {\bf 355}, 845 (2004).


\bibitem{konar_jamr}
[9]C.~Konar, M.~Jamrozy, D.~J.~Saikia, J.~Machalski, MNRAS {\bf 383}, 525 (2008).


\bibitem{jamr_mach}
[10]M.~Jamrozy, J.~Machalski,  K.-H.~Mack, U.~Klein,
   Astron. Astrophys. {\bf 433}, 467 (2005).


\bibitem{jamr_konar}
[11]M.~Jamrozy, C.~Konar, J.~Machalski,  D.~J.~Saikia,
   MNRAS {\bf  383}, 525 (2008).


\bibitem{mach}
[12]J.~Machalski, M.~Jamrozy, S.~Zola, D.~Koziel, Astron. Astrophys.
      {\bf 454}, 85 (2006).


\bibitem{komberg}
[13]  B.~L.~Komberg, I.~N.~Paschenko, Astron. Rep., accepted (2009),
		    arXiv:0901.3721.


\bibitem{verkh_kh}
[14]O.~V.~Verkhodanov, M.~L.~Khabibullina, M.~Singh, et al.,
Eds. Yu.V. Baryshev, I.N.Taganov and P. Teerikorpi, {\bf 1}, 247 (2008).


\bibitem{soboleva}
[15]N.~S.~Soboleva, Astrofiz. Issled., Bull. SAO {\bf 14}, 50 (1981).


\bibitem{nizhelsky}
[16]N.~A.~Nizhelsky, A.~B.~Berlin, A.~M.~Pilipenko et al.,
Abstr. book of Russian astron. conf. VAK-2001, S.Petersburg, p.133 (2001).


\bibitem{baars}
[17]J.~W.~M.~Baars, R.~Genzel,  I.~I.~K.~Pauliny-Toth, A.~Witzel,
 Astron. Astrophys. {\bf 61}, 99 (1977).


\bibitem{aliakb}
[18]K.~D.~Aliakberov,  M.~G.~Mingaliev,  M.~N.~Naugolnaya et al.,
 Astrofiz. issl. (Izv. SAO)
   No {\bf 19}, 60 (1985).


\bibitem{trushk}
[19]S.~A.~Trushkin,
Handbook of the observer at RATAN-600.
{\tt http://w0.sao.ru/hq/lran/manuals/ratan\_manual.html} (2000).


\bibitem{verkh_erukh}
[20]O.~V.~Verkhodanov, B.~L.~Erukhimov, M.~L.~Monosov et al., 
Bull. SAO
{\bf 36},
132
(1993).


\bibitem{verkh}
[21]O.~V.~Verkhodanov,
  in {\it ``Astronomical Data Analysis Software and Systems VI''},
    eds. G.Hunt \& H.E.Payne, ASP Conf. Ser., {\bf 125}, 46 (1997).


\bibitem{condon_cotton}
[22]J.~J.~Condon, W.~D.~Cotton,  E.~W.~Greisen, et al.,
 Astron. J. {\bf 115}, 1693 (1998).


\bibitem{rengel}
[23]R.~B.~Rengelink,  Y.~Tang, A.~G.~de Bruyn,  et al.,
   Astron. Astrophys. Suppl. Ser. {\bf 124}, 259 (1997)


\bibitem{greg}
[24]P.~C.~Gregory, W.~K.~Scott,  K.~Douglas,   J.~J.~Condon,
    Astrophys. J. Suppl. {\bf 103},  427 (1996)


\bibitem{verkh_trushk}
[25]O.~V.~Verkhodanov, S.~A.~Trushkin, H.~Andernach and V.~N.~Chernenkov,
    Bull. SAO {\bf 58}, 118 (2005), arXiv:0705.2959.


\bibitem{verkh2}
[26] O.~V.~Verkhodanov,
    Proc. XXVII radio astron. conf.
    ``Problems of the modern radio astronomy''
    (S.-Petersburg: Inst.Apl.Astron. RAS, 1997, v.1, v.322).

\end{thebibliography}
\end{document}